\documentclass[12pt]{iopart}
\usepackage{graphicx,color}
\usepackage{hyperref}
\usepackage[square,sort&compress,numbers]{natbib}
\usepackage{booktabs}

\usepackage{epsfig}
\usepackage{iopams}

\usepackage{iopams}  

\begin{document}

\title{On the star formation efficiency in high redshift Lyman-$\alpha$ emitters}

\author{Arnab Sarkar}

\address{Department of Physics and Astronomy, University of Kentucky, KY-40508, USA.}
\address{Department of Physics, Presidency University, 86/1 College street, Kolkata, 700073, INDIA}
\ead{arnab.sarkar@uky.edu}

\author{Saumyadip Samui}
\address{Department of Physics, Presidency University, 86/1 College street, Kolkata, 700073, INDIA}
\ead{saumyadip.physics@presiuniv.ac.in}

\begin{abstract}
We present semi-analytical models of high redshift Lyman-$\alpha$ emitters (LAEs) in order to
constrain the star formation efficiency in those galaxies. Our supernova feedback
induced star formation model along with Sheth-Tormman halo mass function correctly
reproduces the shape, amplitude and the redshift evolution of UV and Lyman-$\alpha$
luminosity functions of LAEs in the redshift range $z=2$ to 7.3. We show that the fraction of Lyman-$\alpha$ emitting
galaxies increases with increasing redshifts reaching to unity just after the reionisation.
However, we show that star formation efficiency in those LAEs does not show
any redshift evolution within the uncertainty in available observations. This would
have significant repercussion on the reionisation of the intergalactic
medium.
\end{abstract}
\noindent{\it Galaxies: high-redshift, intergalactic medium, stars: supernovae: general \/}

\maketitle
\section{Introduction}

High redshift galaxies with strong Lyman-$\alpha$ emission (i.e. Lyman-$\alpha$ emitters) 
are detected using narrow band searches with targeted redshifts. 
Such narrow band technique is successful in detecting Lyman-$\alpha$ emitters (LAEs)
in the redshift range $2\:\leq\:z\:\leq\:7.3$ \citep[i.e.][]{2005PASJ...57..165T,
2006ApJ...648....7K,2006PASJ...58..313S,2007ApJ...667...79G,2007ApJS..172..523M,
2007ApJ...671.1227D,2008ApJS..176..301O,2010ApJ...723..869O,2012ApJ...744..110C,
2013MNRAS.431.3589Z,2014ApJ...797...16K,2017MNRAS.466.1242S,2018MNRAS.477.2817S,2018PASJ...70S..13O}. However, their detectability depends
on the emissivity of Lyman-$\alpha$ line and the radiative transport of it
through the galaxy as well as through the intergalactic medium (IGM).
In addition to narrow band searches galaxies are regularly identified using
``drop-out" technique \citep[i.e.][]{2003ApJ...592..728S} even upto a redshift
$z\:\sim 10$ \citep{2004ApJ...616L..79B,2006ApJ...651..142H,2006A&A...456..861R,2015ApJ...803...34B,2018ApJ...855..105O}.
Galaxies detected by this technique are known as Lyman break galaxies (LBGs).
{
Unlike narrow band searches, the dropout technique is 
very efficient in selecting galaxies with strong stellar UV continuum
and hence biased by the UV luminosity of galaxies.
}
Thus, these two techniques are successful in
detecting galaxies with different selection bias and provide useful 
constraints on different physical properties of such galaxies.
Therefore, one should consider both of them together in any theoretical
study of galaxy evolution.
Even though present day improved observational technology has made an
impressive number of observational studies of LAEs available, the theoretical
understanding of them is still in preliminary stages. This is because,
{
the Lyman-$\alpha$ line is a resonant transition and
}
the Lyman-$\alpha$ emissivity depends on many physical properties of the host
galaxy such as amount of star formation, initial mass function of stars,
the dust and neutral hydrogen content and the velocity field of the interstellar
medium that governs the Lyman-$\alpha$ escape fraction, the duty cycle of
Lyman-$\alpha$ phase etc. These are still poorly constraint from the present day
observations.
Several studies of Lyman-$\alpha$ emitters are available in the literature using
simulation \citep{2004ApJ...604L...1B,2006MNRAS.370..273D,2006ApJ...645..792T,2007MNRAS.380L..49S,2009ApJ...696..853L,2011ApJ...726...38Z,2017ApJ...839...44S,2018PASJ...70...55I}  and semi-analytical models   \citep{2007ApJ...670..919K,2008MNRAS.389.1683D,2009MNRAS.398.2061S,1999ApJ...518..138H,2007MNRAS.379..253D,2007ApJ...667..655M,2007ApJ...668..627S,2008MNRAS.384.1363F,2007MNRAS.379..253D}.

However, recent advancement of observations extending to higher redshifts, especially the constraints on the escape fraction of Lyman-$\alpha$ photons in high redshift galaxies \citep[i.e.][]{2011ApJ...730....8H}
has enabled us to revisit it again. Further, it is well demonstrated that the supernova
feedback is very important in determining the star formation in galaxies even in high redshifts
\citep[i.e.][]{2018arXiv180505945S}. Therefore, in this work, we explore the luminosity functions
of high redshift Lyman-$\alpha$ emitters taking into the supernova feedback in the star formation\citep{2003MNRAS.339..289S,2014MNRAS.445.2545D,2014NewA...30...89S,2017MNRAS.472.1576F}.

Further note that star formation efficiency (SFE) of high redshift galaxies
is a very important physical property as it plays the key role in
any processes associated with galaxy evolution\citep[][]{2016MNRAS.460..417S}.
{
Especially it decides
galaxy driven processes such as reionisation history}
and Lyman-$\alpha$ emitting galaxies are
likely to play important role in reionisation.
The star formation efficiency is basically defined as the fraction of baryonic gas
that is converted to stars in a virialized dark matter halo.
In any semi-analytical model it is the most basic parameter that one assumes.
Thus it is important to constrain this from available observations.
Some previous works such as \citep{2009ApJ...704..724T}, using cosmological simulation,
claimed that SFE around $2.5\%$ is enough to fit the observed Lyman-$\alpha$ luminosity function
at $z=3.1$.
Further, \citep{2007MNRAS.379..253D} found a non evolving nature in SFE and suggesting
the SFE be 10\% between $z=5.7$ and 6.5. By considering early
reionization model(ERM) and late reionization model(LRM),
\citep{2008MNRAS.389.1683D} also provided a constant SFE value for $z=4.5$ to
6.56 ( 10\% for ERM and 8\% for LRM).
In this work we also constrain the star formation efficiency in Lyman-$\alpha$ emitting galaxies
in a wide redshift range of $2\le z \le 7.3$ using all updated observations.

The paper is organised as follows. In next section we briefly describe our semi-analytical
model. Our results are discussed in section~3 and finally in section~4 we conclude.
Here, we consider the $\Lambda$CDM cosmology frame work and use the cosmological parameters
of $WMAP5$ data \citep[e.g.,][]{2009ApJS..180..306D} ($\Omega_\circ = 1$, $\Omega_{m} = 0.26$,
$\Omega_{\Lambda} = 0.74$, $\Omega_{b} = 0.044$ and $h = 0.71$). 

\section{Lyman- \texorpdfstring{$\alpha$}{Lg} Luminosity Function}
We first proceed to estimate Lyman-$\alpha$ luminosity functions of galaxies
at different redshifts.
In order to do so, we consider the supernova
{ feedback regulated star formation model}
of \citep{2014NewA...30...89S} for individual galaxies.
We briefly describe the model here.

\subsection{Star formation model}
The baryonic gas inside a dark matter halo, after virialization, is heated up to
the virial temperature of the halo. The gas then cools down due to radiative
cooling and accretes to the centre of the halo. Such accretion of cold
baryonic gas towards the centre of halo enhances the baryonic density
at the central part and leads to star formation in a galaxy \citep{2018PhR...780....1D}. 

On the other hand, massive stars explode as supernovae in relatively short time
scale, which drives the cold baryonic gas out of the galaxy. Such outflow
reduces the star formation in the host galaxy as availability of cold baryonic
gas goes down. We assume that the outflowing mass is proportional
to the instantaneous star formation and the star formation is proportional
to the available cold gas.
{
Such an assumption is motivated by Kennicutt-Schmidt law \citep{1998ApJ...498..541K}
that says the star formation rate is proportional to gas density.}
Finally, the star formation rate ($\dot{M}_{*}$)
in a halo
of total mass $M$ evolves as \citep{2014NewA...30...89S}
\begin{equation}
\dot{M_*}=\frac{M_bf_*}{\kappa\tau\eta_w}[e^{-\frac{t}{\kappa\tau}}-e^{-(1+\eta_w)\frac{t}{\kappa\tau}}].
	\label{eqn_sfr}
\end{equation}
Here, $f_*$ governs the star formation efficiency of the galaxy and $\kappa$
determines the duration of star formation activity in terms of the dynamical time scale,
$\tau$. Throughout this work we have assumed $\kappa$ = 4, which is determined by constraining
UV lumniosity functions of LBGs \citep{2018arXiv180505945S}.
Further, $M_{b}$ is the total baryonic gas mass in the halo which is
taken to be $M_b=(\Omega_b/\Omega_m)M$. The supernova feedback process
is regulated by the parameter $\eta_{w}$, defined as $\dot{M_w} = 
\eta_{w}\dot{M_*}$ \citep{2014NewA...30...89S}, where $M_w$ is the baryonic
mass driven out from the host galaxy by the outflow and over dot represents
the time derivative.
Note that, depending on the outflow mechanism, $\eta_w$ can be related
to the circular velocity of the galaxy ($v_c$) as $\eta_w=
(v_{c}^{0}/v_{c})^{\alpha}$ \citep[e.g.,][]{1977ApJ...218..377W,
1988RvMP...60....1O,2002ApJ...574..590S,2005ARA&A..43..769V,
2008MNRAS.385..783S}; $v_{c}^{0}$ is the circular velocity for the galaxy where
$\eta_{w}=1$.
Further, if the outflow is driven by the hot gas and/or cosmic rays
produced in supernovae shocks, then $\alpha=2$. On the other hand, if
the momentum of the gas drives the outflow, \citep{2017MNRAS.472.1576F,2018PhR...780....1D} $\alpha = 1$. It was shown
by \citep{2014NewA...30...89S} \citep[also see][]{2018arXiv180505945S} that $\alpha=2$ model
is preferred by various observations of high redshift galaxies and hence
we use it here along with $v_{c}^{0}=100$~km/s.

Note that, the baryonic gas in halos collapsed in the neutral region
of the universe can
cool in presence of atomic hydrogen and host star formation, if the
virial temperature ($T_v$) of the halo is greater than $10^{4}$~K. Below this
temperature (and hence in halos with $T_v<10^4$~K) gas can cool only in presence of molecular hydrogen.
In this work, we only consider galaxies that are cooled via atomic hydrogen
cooling.
{
This leads to a minimum halo mass of $2.5\times10^8$~M$_\odot$ that can host star formation
at $z=9$. 
}
Further because of radiative feedback, galaxies collapsed in ionized region of
the universe, due to the 
increased in the Jean's mass, can host star formation if the circular velocity
is $\geq 35$~km/s \citep[see][]{2002ApJ...575..111B,2002MNRAS.333..156B,
2004ApJ...601..666D,2014NewA...30...89S}. For this we assume a complete
suppression of star formation in galaxies with $v_c\le35$~km/s and
no suppression in galaxies with  $v_c\ge110$~km/s.
In the intermediate halo mass region,
{ i.e. for halo mass with 35~km/s$~\le v_c\le 110$~km/s , we have used a linear suppression factor
from $0$ to $1$ by which the star formation is reduced in such halos.
}
Further, AGNs activities
in high mass galaxies are likely to produce a negative feedback on
star formation in those galaxies \citep{2006MNRAS.370..645B,2006MNRAS.368L..67B}. In order to model that, we also consider
a suppression factor of $[1+(M/10^2M_{\odot})^3]^{-1}$ on star formation
in high mass halos due to possible AGN feedback.
{ Such a scenario explains the bright end of the UV luminosity functions
of LBGs \citep{2018arXiv180505945S}}. 

\subsection{Luminosity functions}
The star formation described above will produce stars of different masses
(we assume a Salpeter initial mass function of stars in the mass range
$1-100$~M$_\odot$). The UV photons coming from massive stars can ionize neutral
hydrogen of interstellar medium (ISM). Recombination of those ionised
hydrogen can lead to production of Lyman-$\alpha$ photons. In a case~B recombination scenario,
$\sim 2/3$ of ionising photons produce Lyman-$\alpha$ photons \citep{1989NYASA.571...99O}.
Thus, the star formation rate (i.e. Eq.~\ref{eqn_sfr}) can be used to calculate
the Lyman-$\alpha$ luminosity($L_{Ly\alpha}^{int}$) produced in a star forming galaxies,
i.e. \citep{2009MNRAS.398.2061S},
\begin{equation}
L_{Ly\alpha}^{int}=0.68\:h\nu _\alpha\:(1-f_{esc})\:N_\gamma\:\dot{M_*}.
	\label{eqn_lyain}
\end{equation}
Here, $h\nu_{\alpha}$ is the energy of a Lyman-$\alpha$ photon, i.e.
$h\nu_{\alpha} = 10.2$~eV and $f_{esc}$ is the escape fraction
of the hydrogen ionizing photons from the host galaxy. Further, $N_{\gamma}$
represents number of hydrogen ionizing photons produced per unit baryonic mass
of star formation and it depends on the initial mass function and the metalicity
of the gas. For our work we have taken $N_{\gamma} = 10,840$ per baryonic mass
\citep{1999ApJS..123....3L,2007MNRAS.377..285S}.
Escape fraction of ionizing photon ($f_{esc}$) from host galaxy
is a poorly known quantity from observation \citep{2009ApJ...704..724T,2006ApJ...651..688S,2009ApJ...692.1287I}.
In our work we have used $f_{esc} = 0.1$\citep{2018IAUS..333..254G} that self consistently reproduces
the observational constraints on reionisation. We will also consider
how our model predictions differ for a range of $f_{esc}$ as it is constrained
from observations\citep{2017MNRAS.465.3637M,2018A&A...616A..30C}.

Note that Eq.~\ref{eqn_lyain} provides the intrinsic Lyman-$\alpha$
luminosity of a galaxy. However, Lyman-$\alpha$ luminosity that we
observe is less than that because it can be absorbed in the
host galaxy ISM as well as in the IGM by the dust as well as neutral
hydrogen. We consider a fraction $f^{Ly\alpha}_{esc}$ of the total
Lyman-$\alpha$ finally reaches to us. Thus the observed Lyman-$\alpha$
luminosity of a galaxy is given by
\begin{equation}
L^{obs}_{Ly\alpha}=f^{Ly\alpha}_{esc}\:L^{int}_{Ly\alpha}.
\end{equation}

In order to calculate the Lyman-$\alpha$ luminosity functions we need the formation
rate of halos/galaxies at different redshifts. We use the
redshift derivative of Sheth-Tormen (ST) mass function \citep{1999MNRAS.308..119S}
to calculate the formation rate of dark matter halo. Note that redshift
derivative of mass function provides the difference of the formation
and the destruction rate of halos. Here we assume that redshift derivative
of ST mass function closely follows the formation rate of halos
\citep[see][for a detail discussion on it]{2010MNRAS.402.2778S}.
The Lyman-$\alpha$ luminosity function $\Phi(L,z)$ for luminosity $L$
at a given observed redshift $z$ is given by \citep{2014NewA...30...89S}
\begin{equation}
\Phi(L,z)\:dL =\int\limits_{M_{\rm low}}^{\infty}\:\int\limits_z^{\infty}
 dz_{c}~ dM ~ N(M,z_{c}) \:\: \delta[L\: -\: L(M,z,z_{c})]\:dL.
	\label{eqn_phi}
\end{equation}
Here, $N(M,z_{c})$ is the number density
of the dark matter halos having masses between $M$ to $M$+$dM$ and
collapsed between $z_{c}$ and $z_{c}+dz_{c}$, obtained from the ST mass
function. The delta function $\delta[L\: -\: L(M,z,z_{c})]$, ensures that
the integral survives only for those galaxies with mass $M$
which formed at $z_{c}$ greater than the observed redshift $z$
and having observed Lyman-$\alpha$ luminosity, $L(M,\:z,\:z_{c})$. Further,
the lower limit of the mass integral, $M_{\rm low}$,
is decided by the cooling criteria discussed above.

Note that not all galaxies are likely to show up as Lyman-$\alpha$
emitters. In fact it has been found observationally that only
a fraction of galaxies that are found using the Lyman-break
technique is detectable as Lyman-$\alpha$ emitters  \citep[see][]{2003ApJ...588...65S,2010ApJ...711..693K,2010MNRAS.408.1628S}.
\begin{figure}
\centerline{\includegraphics[height=8cm,angle=0]{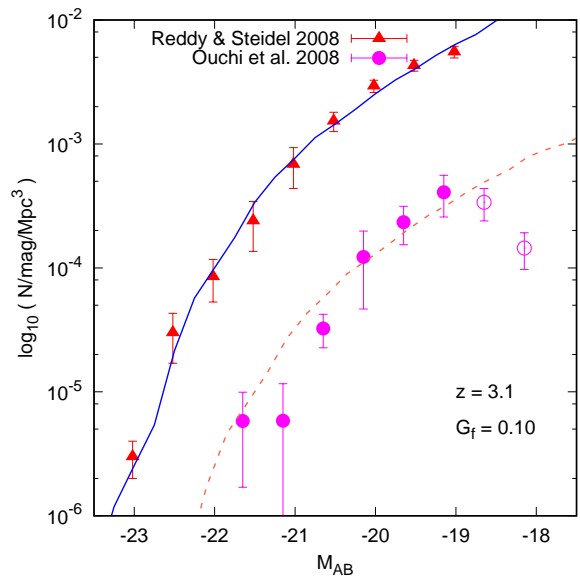}}
 \caption{Above plot shows model predicted UV luminosity function for LBGs
and LAEs at redshift 3.1 along with the obseravtional data. Blue solid line indicates the UV luminosity function for LBGs and red dashed line indicates UV luminosity function for LAEs. Red triangles represent \citep{2009ApJ...692..778R} data of UV LF for Lyman break galaxies and cyan circles (empty and filled) represent the UV LF for Lyman alpha emitters \citep{2008ApJS..176..301O}. Empty cyan circles, due to incompleteness in observational data, are excluded from fitting process.
}
\label{fig:image1}
\end{figure}
Thus, in our work we assume that a fraction, $G_f$ of all galaxies
shows up as Lyman-$\alpha$ emitters. We simultaneously fit observed
UV luminosity function of Lyman-break galaxies (LBGs) and
UV luminosity function of Lyman-$\alpha$ selected sample at a similar redshift
to obtained the value of $G_f$. In order to obtained UV luminosity functions
of LBGs we follow \citep{2014NewA...30...89S}. Note that Eq.~\ref{eqn_phi} can be
used to calculate UV luminosity function of LBGs if the luminosity
$L$ is the UV luminosity of the galaxy that can be obtained by convolving
the star formation rate with UV luminosity of a single burst of star
formation \citep[see][for detail]{2007MNRAS.377..285S}. { Further, the UV luminosity
is also affected by the dust in the galaxies i.e. dust attenuation and like the Lyman-$\alpha$
luminosity we assume that only a fraction $1/\eta$ of the intrinsic UV luminosity
is finally reached to us.
Thus, we use the combination, $f_*/\eta$
as a free parameter of our model and fit the observed UV luminosity
function of LBGs at different redshifts by varying that \citep{2007MNRAS.377..285S}.
}
On the other hand we vary $f_*f^{Ly\alpha}_{esc}$ combination with redshifts
to fit observed Lyman-$\alpha$ luminosity functions at different redshifts.



\section{Results}
In this section we show our model predictions and compare them with the available
observations in order to constrain our model parameters.
We first concentrate at redshift $z=3$. In Fig.~\ref{fig:image1}
we have shown the UV luminosity functions of LBGs at $z=3$ as predicted by our
model by the solid red line. The corresponding observational data are shown
by the red filled triangles with error bars adopted from \citep{2009ApJ...692..778R}.
We have used $\chi^2$ minimization technique to match our model with observation.
It is clear from the figure that a reasonably good agreement is obtained
between the model and observed data with value of $f_*/\eta=0.144$. 
We now turn to the UV luminosity function of Lyman-$\alpha$ emitters
in the similar redshift, i.e. $z=3.1$. As mentioned above we assume that
a fraction $G_f$ of all LBGs shows up as LAEs. Thus we scaled the above fitted UV
luminosity function of LBGs by factor $G_f$ to match with the
observational UV luminosity function of LAEs keeping all other parameters the same.
\begin{figure}[h!!!!!!!]
\centerline{\includegraphics[height=8cm,angle=0]{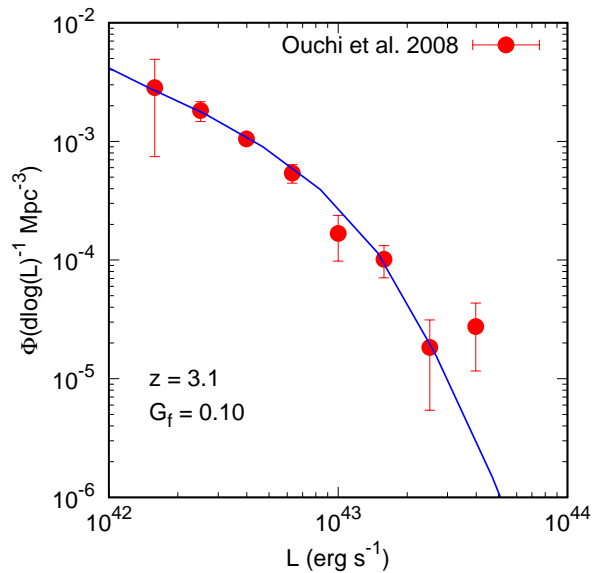}}
\caption{Lyman-$\alpha$ luminosity function of Lyman-$\alpha$ emitters at $z=3.1$. The solid line is for our best fit model and the data points are taken from \citep{2008ApJS..176..301O}.}
	\label{fig:LAz3.1}
\end{figure}
The model prediction and the observational
data from \citep{2008ApJS..176..301O} are also shown in Fig.~\ref{fig:image1}
by the dashed red curve and solid magenta filled circles with error bars
respectively. We see that a good agreement is obtained with $G_f=0.10$.
Similar values were obtained by \citep{2003ApJ...588...65S} and \citep{2008MNRAS.385..783S}.
Thus we conclude that only 10\% of total galaxies shows up
as LAEs at $z=3.1$. Note that, due to larger uncertainties in the
UV luminosity function of LAEs we do not use $\chi^2$-mechanism to obtain the value
of $G_f$. Further, due to incompleteness in the observed data points
of the two lowest luminosity bins (as shown by open circles) we
omitted them from fitting \citep[see][for details]{2008ApJS..176..301O}.
This $G_f$ has been used in finding the Lyman-$\alpha$ luminosity functions of Lyman-$\alpha$ emitters
that we consider next.

\begin{figure}
\begin{center}
	\resizebox{5.1cm}{!}{\includegraphics[angle=0]{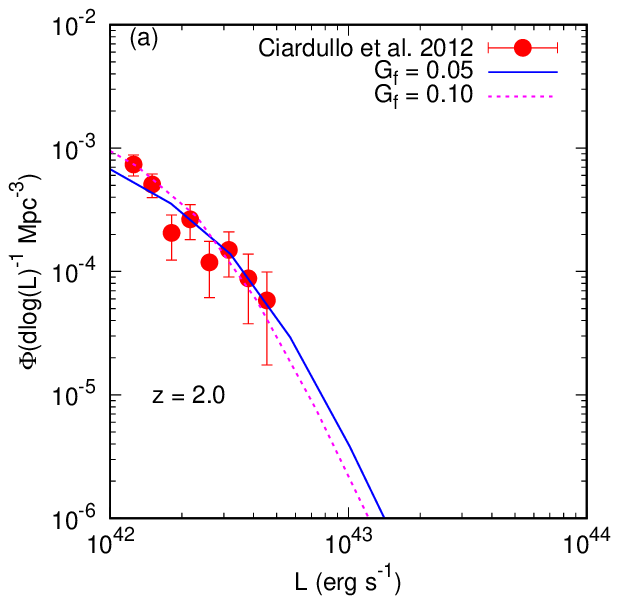}}
	\resizebox{5.1cm}{!}{\includegraphics[angle=0]{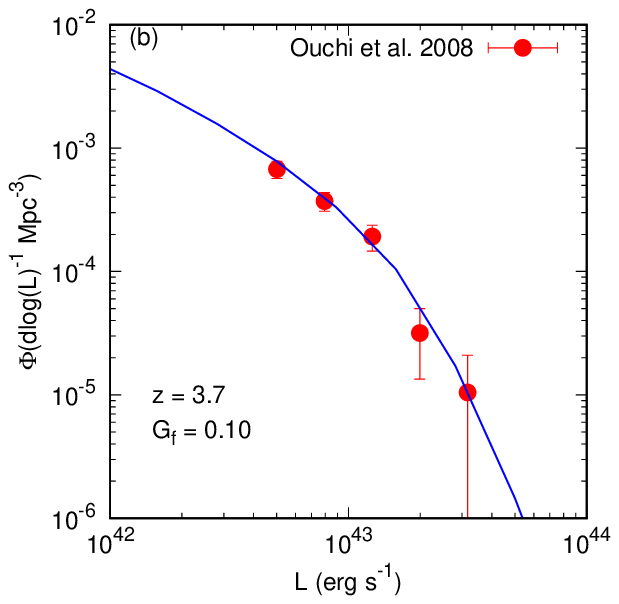}}
	\resizebox{5.1cm}{!}{\includegraphics[angle=0]{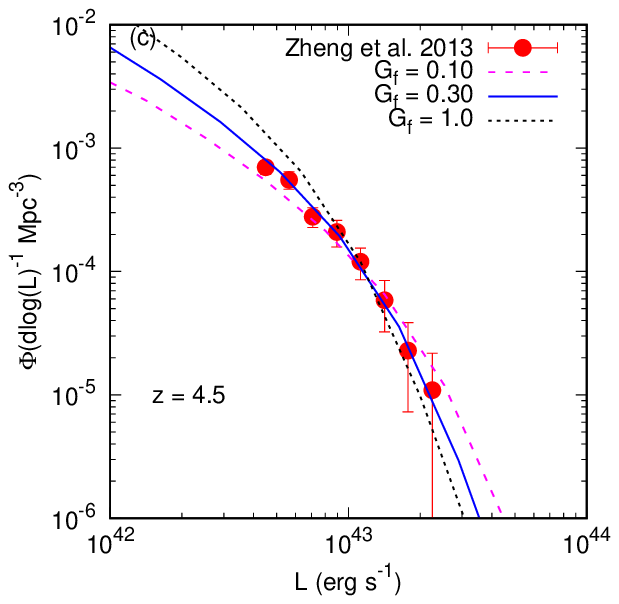}}
	\resizebox{5.1cm}{!}{\includegraphics[angle=0]{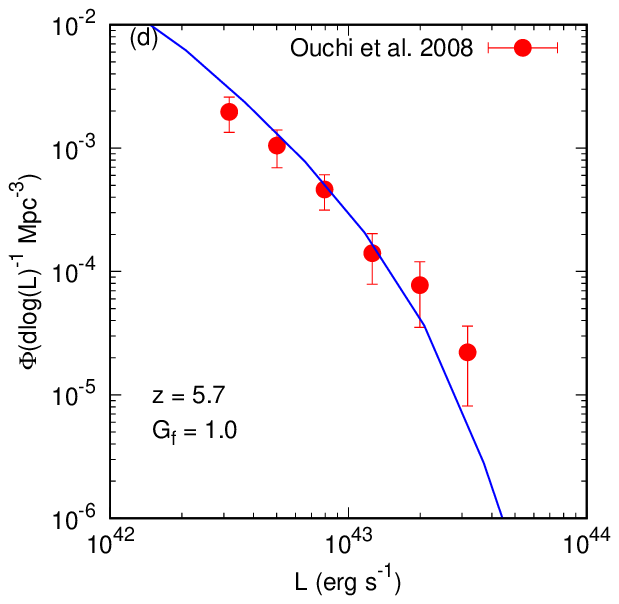}}
	\resizebox{5.1cm}{!}{\includegraphics[angle=0]{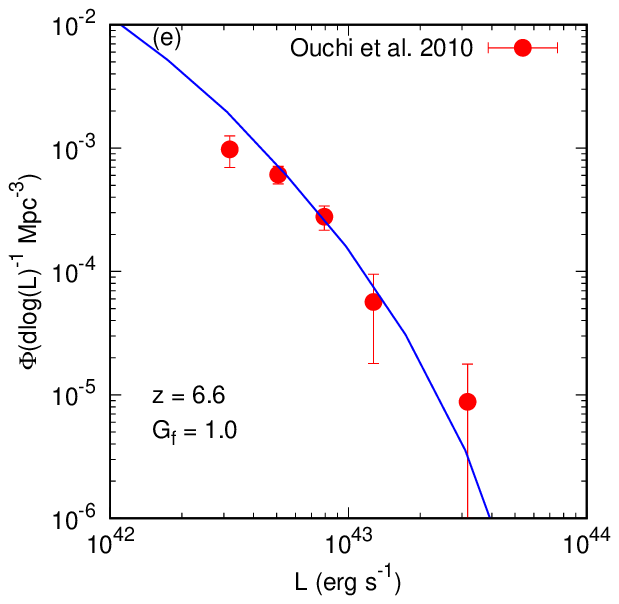}}
	\resizebox{5.1cm}{!}{\includegraphics[angle=0]{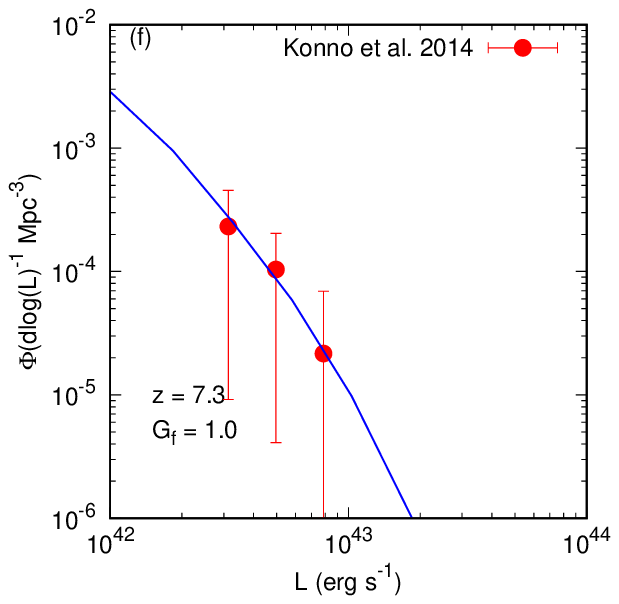}}
%
\caption{Above plots are for Lyman-$\alpha$ luminosity function for LAEs as predicted by our model
for redshift 2.0, 3.7, 4.5, 5.7, 6.6, 7.3 along with the observational data. In all panels,
blue solid line indicates our model predicted luminosity functions.
Red filled circles represent the observational data of Lyman-$\alpha$ luminosity function taken from
\citep{2012ApJ...744..110C} ($z=2$), \citep{2008ApJS..176..301O}($z=3.7$ and 5.7),
\citep{2013MNRAS.431.3589Z}($z=4.5$), \citep{2010ApJ...723..869O}($z=6.6$) and \citep{2014ApJ...797...16K}($z=7.3$).
The $G_f$ factors for each redshift are indicated in each panel.
In panel~(a) we also show our model prediction for $G_f=0.10$ with dot-dashed line.
For $z=4.5$ (panel~c) the model predictions with $G_f=0.10$ and 1.0 are shown
by dashed and dotted-dashed curves respectively.
}
\label{fig:image2}
\end{center}
\end{figure} 

\begin{table}
  \centering{
  \caption{Comparison of different best fit model parameters for different redshifts.}}
  \label{tab:table1}
  \begin{tabular}{cccccccc}
    \toprule
	  $z$ & ${f^{Ly\alpha}_{esc}}^\dagger$ & $G_f$ &  $f_*\:f^{Ly\alpha}_{esc}$ & $\chi^2/d.o.f$ & $f_*$ & $\frac {M_*}{M_b}^{\ddagger}_\S$  \\
    \midrule

2.0 & $0.05\:\pm\:0.04$	& 0.05 & $0.018\:\pm\:0.004$ & 0.83 & $0.32\:\pm\:0.24$ & $0.17\:\pm\:0.13$ \\
	& 				& 0.10 & $0.013\:\pm\:0.002$ & 1.04 & $0.26\:\pm\:0.20$ & $0.13\:\pm\:0.10$ \\
\midrule
3.1	& $0.07\:\pm\:0.05$ & 0.10 & $0.047\:\pm\:0.007$ & 0.52 & $0.67\:\pm\:0.45$ & $0.37\:\pm\:0.24$ \\
\midrule
3.7   & $0.05\:\pm\:0.02$ & 0.10 & $0.050\:\pm\:0.016$ & 0.46 & $1.0\:\pm\:0.5$ & $0.56\:\pm\:0.28$ \\
\midrule
4.5 & $0.13\:\pm\:0.07$ & 0.10 & $0.045\:\pm\:0.012$ & 1.87 & $0.34\:\pm\:0.20$ & $0.21\:\pm\:0.12$ \\
	& 				& 0.30 & $0.029\:\pm\:0.008$ & 0.82 & $0.22\:\pm\:0.13$  & $0.13\:\pm\:0.08$ \\
	&				& 1.0   & $0.020\:\pm\:0.004$ & 1.40 & $0.15\:\pm\:0.07$ & $0.10\:\pm\:0.04$ \\
\midrule
5.7 & $0.36\:\pm\:0.20$ & 1.0 & $0.037\:\pm\:0.013$ & 2.07 & $0.10\:\pm\:0.06$ & $0.06\:\pm\:0.04$\\
\midrule
6.6 &$0.33\:\pm\:0.15$ & 1.0 & $0.055\:\pm\:0.013$ & 2.00  & $0.15\:\pm\:0.09$ & $0.10\:\pm\:0.06$ \\
\midrule
7.3 & $0.09\:\pm\:0.11$ & 1.0 & $0.058\:\pm\:0.012$ & 0.72 & $0.64\:\pm\:0.78$  & $0.49\:\pm\:0.55$ \\
 
    \bottomrule  
\end{tabular}
	\begin{tabular}{l}
		$^\dagger${\footnotesize{
			Escape fraction of Lyman-$\alpha$ photons of Lyman-$\alpha$ emitters. Data are taken from \citep{2011ApJ...730....8H}.}}\\
		$_\S${\footnotesize{ Ratio of star mass to baryonic mass of Lyman-$\alpha$ emitting galaxies as determined by $\frac{f_*}{1+\eta_w}$ \citep{2014NewA...30...89S} }}\\
                $_\ddagger${\footnotesize{$\eta_w$ is calculated for $M=10^{11}M_\odot$}}
	\end{tabular}
\end{table}
Fig.~\ref{fig:LAz3.1} shows our model prediction of Lyman-$\alpha$ luminosity function at $z=3.1$ by solid line and the observed data points from \citep{2008ApJS..176..301O} (filled circles). Here also, we use
$\chi^2$-minimization to fit the model with observation by varying $f_*\:f^{Ly\alpha}_{esc}$.
We can see that our model well reproduces the shape and amplitude of the Lyman-$\alpha$
luminosity function of LAEs.
Thus our feedback induced star formation model provides a good description of
Lyman-$\alpha$ emitters at $z=3.1$.
The fitted value along with $1\sigma$ uncertainty for
$f_*\:f^{Ly\alpha}_{esc}$ is  $0.047 \pm 0.007$ at $z=3.1$.
Now, as already mentioned \citep{2011ApJ...730....8H} has measured
$f^{Ly\alpha}_{esc}=0.07\pm 0.04$ from observations at $z=3$. Using
this we estimate $f_*=0.67 \pm 0.45$ where we add the observational uncertainty
in $f^{Ly\alpha}_{esc}$
and the fitting uncertainty of $f_*\:f^{Ly\alpha}_{esc}$ in quadrature.
{
Note that the bright end of the observed Lyman-$\alpha$ luminosity
function suffer from cosmic variance due to limited survey volume and
future large volume survey will help us to understand the nature of such
bright LAEs.
}

The procedure described above for $z\sim 3$ has been followed for all other
redshift bins, $z=2.0,$ 3.7, 4.5, 5.7, 6.6 and 7.3. The resulting values of
$G_f$, $f_*\:f^{Ly\alpha}_{esc}$ and hence the $f_*$ are tabulated
in Table~\ref{tab:table1}. We also provide the best fit $\chi^2$ values.
The fitted Lyman-$\alpha$ luminosity functions
of LAEs and the observational data are shown in Fig.~\ref{fig:image2}. Note that the observed
UV luminosity function of LAEs in all these redshifts are not available
except for $z=3.1$, 3.7 and 5.7 and hence $G_f$ can not be estimated
at those redshifts (i.e. at $z=2$, 4.5, 6.6 and 7.3) using the procedure describe earlier for $z=3$.
In absence of this, at $z=2.0$ we have
used $G_f=0.10$ as obtained from near by redshift (i.e. at $z=3.1$)
as well as $G_f=0.05$ because of the trend seen from other redshifts
that $G_f$ decreases with decreasing redshift. From the value of $\chi^2$
we see that both these values provide similar fit the observed Lyman-$\alpha$
luminosity function.
For $z=4.5$, using
$G_f$ as obtained at $z=3.7$ or $5.7$ our model predictions
do not provide a good fit to the observational data as can be seen from
the value of $\chi^2$. An intermediate value $G_f=0.30$ provide the best
fit in this redshifts. This is also consistent with the increasing trend of
$G_f$ with increasing redshift. For $z>5.7$ we have used $G_f=1.0$ as obtained
at $z=5.7$, and we get good fit of the model predictions with observational
data. Thus we conclude that even though the fraction of galaxies, that
are detected through narrow band Lyman-$\alpha$ emission is only
$\lsim 10$ \% at $z=2$, it increases rapidly with increasing redshift
and reaches to unity just after the end of the reionisation process.
Hence, during the reionisaton period all galaxies are expected to have
strong Lyman-$\alpha$ emission. Similar results were
obtained by \citep{2008MNRAS.385..783S} in spite of the fact that
they did not take account of supernova feedback in star formation. Further, it is clear from the Fig.~\ref{fig:image2} and the $\chi^2$ per degrees of
freedom as given in Table~\ref{tab:table1} that our models provide
a good fit to the observed Lyman-$\alpha$ luminosity function of LAEs in the entire
redshift range from $z=7.3$ to 2.0. The best fit values of
$f_*f_{esc}^{Ly\alpha}$ along with $1-\sigma$ uncertainty are reported
in column~4 of Table~\ref{tab:table1}.
Thus we can say that the  SNe
feedback is operating in the high redshift galaxies with strong
Lyman-$\alpha$ emission. 

\begin{figure}[h!!!!!!!!!!!!!!!!!!!!!!!!!]
\centerline{\includegraphics[height=7cm,angle=0]{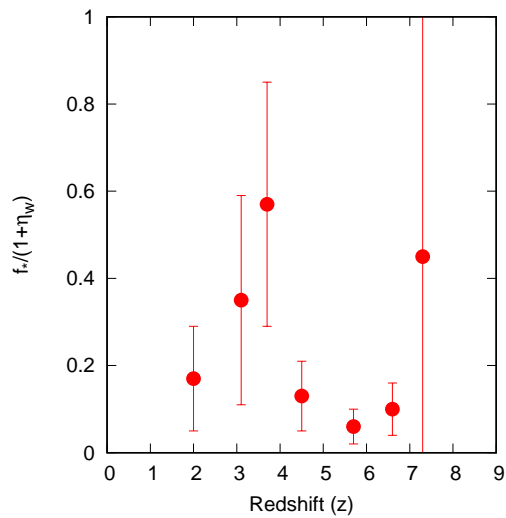}}
\caption{Above plot represents the star formation efficiency (SFE) with redshift. Red filled circles denotes the SFE for each redshift bin.}
\label{fig:image3}
\end{figure}

Most interesting result of our work is the redshift evolution
of star formation efficiency in LAEs as can be seen from the
values of $f_*$ at different redshifts (see Table~\ref{tab:table1}).
Note that the parameter $f_*$ used here in not exactly the
canonically used star formation efficiency. The fraction of
total baryonic gas that will eventually convert to stars in
the galaxy is $f_*/(1+\eta_w)$ in presence of supernova feedback and this should be compared
with the observations of $M_b/M_*$. It is clear from
the Table~\ref{tab:table1} as well as Fig.~\ref{fig:image3}
where we have plotted $f_*/(1+\eta_w)$ as a function of redshift that
within the uncertainty the star formation efficiency does not
show any evolution in the LAEs for the entire redshift range
of $z=2$ to 7.3.
{
Thus, we conclude that even though the fraction
of galaxies that shows up as LAEs changes drastically from
reionisation to late universe, the star formation efficiency
does not change significantly in those Lyman-$\alpha$ emitting galaxies.
In particular, given the observational uncertainty and hence the uncertainty
in derived star formation efficiency, no trend can be identified
in the redshift evolution of star formation efficiency.
}


{
\subsection{Variation of model parameters}
There are two crucial model parameters that are poorly constraint from observations,
namely the escape fraction of UV photon, $f_{esc}$ and the fraction of galaxies
that are detected as LAEs, i.e. $G_f$. Here we show how these two parameters affect
our results.

First we concentrate on $f_{esc}$ that appear in the Eqn.~\ref{eqn_lyain}. Note that this
parameter also regulate the reionisation history and hence we show our results only for
$z=6.6$ i.e. when the IGM is likely to have significant neutral fraction.
Various observations have reported value of $f_{esc}$ in the range of 0.05
to 0.3 \cite{2006ApJ...651..688S,2009ApJ...692.1287I} for high redshift galaxies.
In Fig.~\ref{fig:image4} we show the model predicted Lyman-$\alpha$ luminosity
function along with observed data for $f_{esc}=0.05$, 0.1 and 0.3. It is clear from the
figure that all three models produce similar fit for the observed data. The
$\chi^2$ per d.o.f for these three models are 2.30, 2.00, 2.27 respectively for $f_{esc}=0.05$, 0.1 and 0.3.
The resulting values of star formation efficiency are $f_*= 0.14 \pm 0.07,~~ 0.15 \pm 0.09, ~~ 0.20 \pm 0.10$.
Thus within the uncertainty, the star formation efficiency does not change significantly for a wide
variation in the value of escape fraction of UV photon from the galaxy. We note that similar results
are also obtained from any other redshifts that are being considered in this work.  
\begin{figure}[h!!!!!!!!!!!!!!!!!!!!!!!!!]
\centerline{\includegraphics[height=7cm,angle=0]{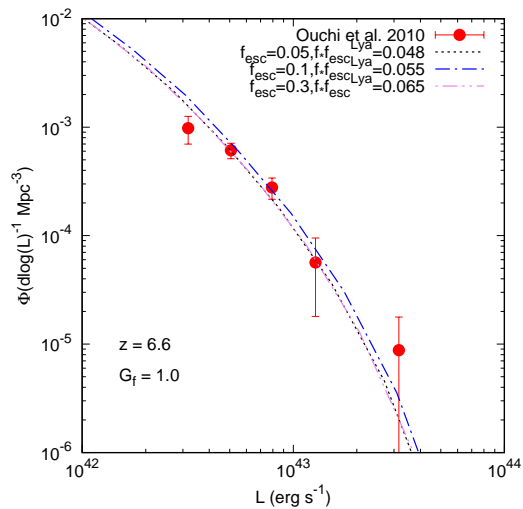}}
\caption{Variation of Lyman-$\alpha$ luminosity function due to variation of $f_{esc}$ at $z=6.6$. The dotted,
dot-dashed and do-dot-dashed curves are for $f_{esc}=0.05$, 0.1 and 0.3 respectively.}
\label{fig:image4}
\end{figure}

Next we consider the variation of $G_f$. Note that recent observations have resulted very low
value of $G_f$ for $z \geq 6$ \citep{2012ApJ...744...83O,2012ApJ...744..179S,2014ApJ...795...20S}.
Thus it is important to see how such variation affect our results. We show in Fig.~\ref{fig:image6}
the Lyman-$\alpha$ luminosity functions at $z=7.3$ for different $G_f=1,$ 0.5 and 0.3. All
three models predict luminosity functions that are consistent with the observation.
Fitted values of minimum $\chi^2$ are unable to distinguish between models given
the large uncertainty in the observed data. However, for lower value of $G_f=0.3$
we need unrealistic value of the star formation efficiency parameter, $f_*=1.0\pm 0.7$.
Thus an improved measurement of the Lyman-$\alpha$ luminosity functions at $z=7.3$
is needed in order to understand the physical processes happening inside these
Lyman-$\alpha$ emitting galaxies.    
\begin{figure}[h!!!!!!!!!!!!!!!!!!!!!!!!!]
\centerline{\includegraphics[height=7cm,angle=0]{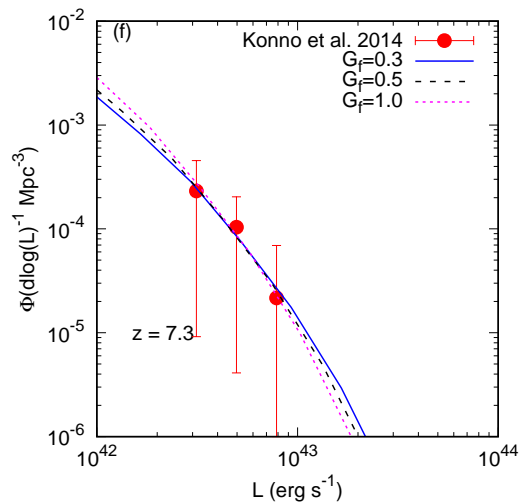}}
\caption{Variation of Lyman-$\alpha$ luminosity function due to variation of $G_f$ at $z=7.3$.}
\label{fig:image6}
\end{figure}
}

\section{Conclusion}

We have presented semi-analytical models of galaxy formation and evolution to understand
the redshift evolution of luminosity functions of LAEs and their physical
properties. In particular we have used star formation model regulated
by supernova feedback along with Sheth-Tormen halo mass function to
obtain simultaneously the UV and Lyman-$\alpha$ luminosity functions of LAEs in the redshift
range $z=2$ to 7.3. Our models correctly reproduce the shape and redshift
evolution of { both UV and Lyman-$\alpha$} luminosity functions of LAEs demonstrating the fact that
the supernova feedback is indeed operational in high redshift Lyman-$\alpha$
emitters.
{
Finally we derive the average star formation efficiency of the Lyman-$\alpha$
emitting galaxies at different redshifts using observational constraint of escape
fraction of Lyman-$\alpha$ emission from galaxies.
}

 We show that the fraction
of Lyman-$\alpha$ emitting galaxies increases with increasing redshift,
reaching to unity just after the end of reionisation, i.e. at $z=5.7$.
On the other hand the star formation efficiency does not vary significantly
before and after reionisation in those Lyman-$\alpha$ emitting galaxies. 
{
This conclusion is independent of the uncertainty in the escape of UV photon from those galaxies.
Such a result was also obtained by previous work by \citep{2007MNRAS.379..253D,2018PASJ...70...55I} who showed that the change in the Lyman-$\alpha$ emitter fraction compared to LBGs
are due to change in the IGM or surrounding halo gas, not due to change in the physical
properties of LAEs. Further, we show that one needs to fit all available data of
LAEs in order to constraint the physical properties of LAEs (also see \citep{2018PASJ...70...55I}.
\citep{2018PASJ...70...55I} also showed that a highly fluctuating $f_{esc}^{Ly\alpha}$ is needed to 
match recent observations that shows very small fraction of LAEs at $z > 6$.   
Our models also require unphysical parameters such as star formation efficiency of order unity
in order to understand such a low fraction of LAEs while fitting simultaneously the UV and
Lyman-$\alpha$ luminosity functions of LAEs.
}
However, we also wish to point out that
this conclusion is highly biased due to large uncertainty in the observed
luminosity functions of LAEs and also in the measurements of the escape
fraction of Lyman-$\alpha$ photons from the galaxy. Precise estimation
of these in a large volume survey would enable us to more accurately
constrain the evolution of the star formation
efficiency in the Lyman-$\alpha$ emitters.

\ack
We thank anonymous referee for suggestions that has improved the paper.  We thank V. Tilvi for providing us new data on current Lyman alpha emitter surveys. AS thanks to R. Ciardullo for his useful comments on different Lyman alpha survey volume. This work is partially funded by the Physics Incentive programme of the University of Kentucky, Kentucky.  SS
thanks Presidency University, Kolkata for providing funds through FRPDF scheme. SS also thanks UGC, India for support through UGC Start Up grant.
    
\providecommand{\newblock}{}

\end{document}